\documentclass[aps,preprint,showpacs,preprintnumbers,amsmath,amssymb]{revtex4}
\usepackage{amsmath,mathrsfs,amsbsy,color,graphicx,bm,amsthm,amsfonts}
\usepackage{units}
\usepackage{bbm}
\usepackage{times}
\usepackage{dcolumn}
\usepackage{mathrsfs}
\usepackage{amsmath,amssymb,epsfig}
\begin{document}

\title{Quantum coherence and distribution of N-partite bosonic fields in noninertial frame }
\author{Shu-Min Wu$^1$, Hao-Sheng Zeng$^2$\footnote{Email: hszeng@hunnu.edu.cn}, Hui-Min Cao$^2$}
\affiliation{$^1$ Department of Physics, Liaoning Normal University, Dalian 116029, China\\
$^2$ Department of Physics, Hunan Normal University, Changsha 410081, China
}


\begin{abstract}
We study the quantum coherence and its distribution of $N$-partite GHZ and W states of bosonic fields in the noninertial frames with arbitrary number of acceleration observers. We find that the coherence of both GHZ and W state reduces with accelerations and freezes in the limit of infinite accelerations. The freezing value of coherence depends on the number of accelerated observers. The coherence of $N$-partite GHZ state is genuinely global and no coherence exists in any subsystems. For the $N$-partite W state, however, the coherence is essentially bipartite types, and the total coherence is equal to the sum of coherence of all the bipartite subsystems.

$\mathbf{Keywords}$: noninertial frame, quantum coherence, N-partite, curved spacetime
\end{abstract}

\vspace*{0.5cm}
\maketitle
\section{Introduction}
Quantum coherence originates from superposition principle of quantum state and is one of the fundamental concepts of quantum physics \cite{Q1}. It has been shown that quantum coherence is a valuable resource in such as quantum thermodynamics,
condensed matter physics, life sciences and quantum information processing tasks \cite{TBMC1,TBMC2,TBMC3,TBMC4,TBMC5,TBMC6,TBMC7,TBMC8,TBMC9,TBMC10,Zeng2019,TBMC11}.
Since Baumgratz $et al.$ put forward the measure of quantum coherence based on resource theory, quantum coherence has been received much attention and research.
In the quantifying methods of quantum coherence, the most intuitive and easily calculated may be the $l_1$ norm of coherence \cite{S11} which is defined as the sum of the absolute value of all the off-diagonal elements of a system density matrix,
\begin{equation}\label{w1}
    C(\rho)=\sum_{{i\neq j}}|\rho_{i,j}|.
\end{equation}
Where the density matrix $[\rho_{ij}]$ is written in a given set of preferred basis $\left\{
{\left| i \right\rangle } \right\}_{i = 1, \ldots ,d}$ of a
$d$-dimensional quantum system, which implies that the definition of quantum coherence is base-dependant. The $l_1$ norm of coherence fulfills the requirements of resource theory and is a distance-based measure of coherence \cite{S11}.

Apart from the pure quantum-mechanical aspect, it is also important to study quantum resource under the relativistic framework \cite{Q12,Adesso2007,Schuller-Mann,Richter2015,Khan2014,Alsing2006,Q13,Q14,Q15,Q16,Q17,S8,S9,Friis2012,Wangjc2020,Dong2020}.
It is not only an interesting theoretical topic from the view of noninertial observers, but also a very practical consideration, because many modern experiments about quantum information involve relativistic photons or particles \cite{AH,AH1}.
In the past years, the effect of Unruh radiation on quantum fields was studied extensively. It was shown that the entanglement of fermionic fields degrades and approaches to a finite value under the infinite acceleration limit \cite{Alsing2006,Q12,Q13,Q15,Q16,Q17,S8,S9}, but the entanglement of bosonic fields vanishes in the infinite acceleration limit \cite{Q14,Adesso2007,Schuller-Mann,Richter2015}. The quantum coherence for multipartite Dirac fields was also studied and the phenomenon of freeze was found \cite{S1}. However, quantum coherence for multipartite bosonic fields is rarely studied, due to its relatively complicated calculations. This makes the motivation of this paper.

In this paper, we firstly discuss the quantum coherence of tripartite systems of bosonic fields in noninertial frame. Assume that Alice, Bob and Charlie share a tripartite GHZ state or W state of bosonic fields in inertial frame, afterward Alice stays in the inertial frame but Bob and Charlie move with uniform accelerations.
We mainly study the influence of the Unruh effect on quantum coherence and its distribution.  Next, we extend the research from tripartite systems to N-particle systems and discuss the relevant problems.

The paper is organized as follows. In Sec.II, the dynamic behaviors of quantum coherence and its distribution for tripartite GHZ and W states in noninertial frame are discussed. In Sec.III, we extend the issues to N-partite systems.  Finally, a summary is arranged in Sec.IV.

\section{Tripartite coherence in noninertial frames}
It is well-known that Minkowski coordinates are suitable to describe quantum fields in inertial frames. For the observer moving with uniform acceleration, however, Rindler coordinates are appropriate for dealing with problems. By solving Klein-Gordon equation and quantizing the free quantum scalar field in the Minkowski and Rindler spacetimes respectively, a transformation between Minkowski and Rindler modes can be established. It has been shown that the Minkowski vacuum from an
inertial observer would be detected as a two-mode squeezed state from the viewpoint of a Rindler observer \cite{Schuller-Mann}
\begin{eqnarray}\label{w2}
|0\rangle_M=\frac{1}{\cosh (s)}\sum_{n=0}^\infty\tanh^n(s)|n\rangle_I|n\rangle_{II},
\end{eqnarray}
and the Minkowski single article state reads
\begin{eqnarray}\label{w3}
|1\rangle_M=\frac{1}{\cosh^2 (s)}\sum_{n=0}^\infty\tanh^n(s)\sqrt{n+1}|n+1\rangle_I|n\rangle_{II}.
\end{eqnarray}
Where $|n\rangle_I$ and $|n\rangle_{II}$ represent respectively the Rindler modes in the causally disconnected regions $I$ and $II$, and the acceleration parameter $s$ is defined by $\sinh(s)=(e^{2\pi\omega/a}-1)^{-\frac{1}{2}}$ with acceleration $a$ and mode frequency $\omega$.

\subsection{Tripartite GHZ state}
Aassume that Alice, Bob and Charlie initially share a tripartite GHZ state of bosonic fields
\cite{S10}
\begin{eqnarray}\label{w4}
|GHZ\rangle_{ABC}=\frac{1}{\sqrt{2}}[|0_{A}0_{B}0_{C}\rangle+|1_{A}1_{B}1_{C}\rangle],
\end{eqnarray}
at the same point in Minkowski spacetime, where the modes $|n_{A}\rangle$, $|n_{B}\rangle$ and $|n_{C}\rangle$ have different frequencies and are detected by Alice, Bob and Charlie respectively.
Then Alice stays in the inertial frame, but Bob and Charlie move with different accelerations. According to Eqs.(\ref{w2}) and (\ref{w3}),  the GHZ state becomes
\begin{eqnarray}\label{w5}
|GHZ\rangle_{AB_IC_IB_{II}C_{II}}&&=\frac{1}{\sqrt{2}}\sum_{m,n=0}^\infty\tanh^m(w)\tanh^n(r)\left[\frac{1}{\cosh w\cosh r}|0mn\rangle|mn\rangle \nonumber\right. \\
&&\left.+\frac{1}{(\cosh w\cosh r)^2}\sqrt{(m+1)(n+1)}|1(m+1)(n+1)\rangle|mn\rangle\right],
\end{eqnarray}
where we abbreviate $|i_Aj_{B_I}k_{C_I}\rangle|l_{B_{II}}p_{C_{II}}\rangle$ as $|ijk\rangle|lp\rangle$,
and use $w$ and $r$ to denote the acceleration parameters for Bob and Charlie, respectively. Obviously, the initial information in region $I$ is leaked partially into the inaccessible region $II$ through accelerating effect.
Tracing over the unaccessible modes $B_{II}$ and $C_{II}$ in region $II$, one obtains the density matrix
\begin{eqnarray}\label{w6}
\rho_{AB_IC_I}&&=\frac{1}{2}\sum_{m,n=0}^\infty\tanh^{2m}w\tanh^{2n}r\left[\frac{1}{\cosh^2 w\cosh^2 r}|0mn\rangle\langle0mn|\right.\nonumber \\
&&+\frac{1}{\cosh^4 w\cosh^4 r}(m+1)(n+1)|1(m+1)(n+1)\rangle\langle1(m+1)(n+1)| \\
\nonumber &&\left.+\frac{1}{\cosh^3 w\cosh^3 r}\sqrt{(m+1)(n+1)}\{|0mn\rangle\langle1(m+1)(n+1)|+h.c.\}\right].
\end{eqnarray}
According to Eq.(\ref{w1}), the coherence of this state can be expressed as
\begin{eqnarray}\label{w7}
 C(\rho_{AB_IC_I})=\frac{1}{(\cosh w \cosh r)^{3} } (\sum_{m=0}^\infty\sqrt{m+1}\tanh^{2m}w)(\sum_{n=0}^\infty\sqrt{n+1}\tanh^{2n}r).
\end{eqnarray}
We can see that quantum coherence $C(\rho_{AB_IC_I})$ depends on
acceleration parameters $w$ and $r$, meaning that the thermal noise introduced by the Unruh radiation influences quantum coherence.

\begin{figure}
\includegraphics[scale=1.0]{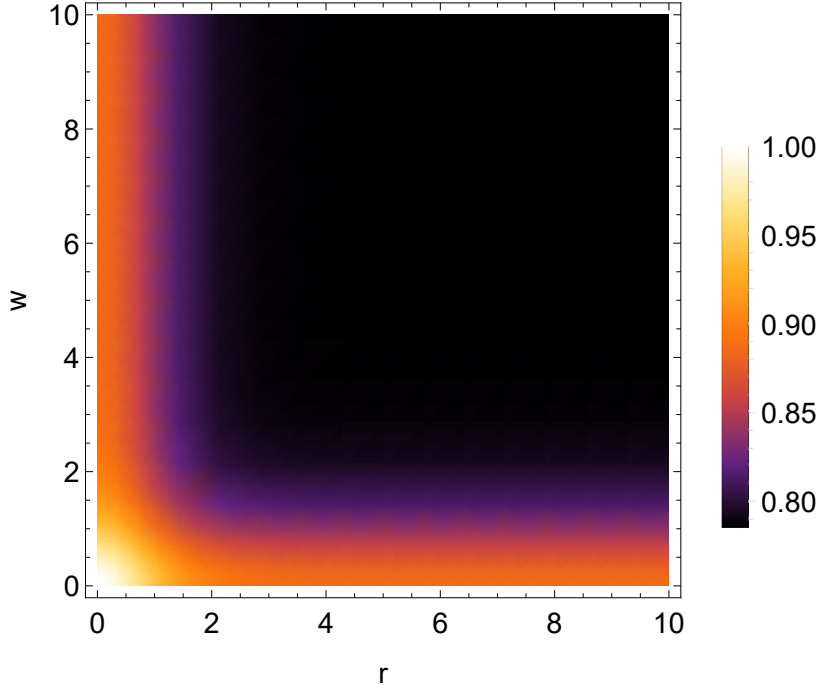}
\caption{Quantum coherence $C(\rho_{AB_IC_I})$ of the GHZ state as a function of  the acceleration parameters $r$ and $w$.
 }
\label{Fig1}
\end{figure}

In Fig.\ref{Fig1}, we plot the coherence $C(\rho_{AB_IC_I})$ of the GHZ state as a function of the acceleration parameters $r$ and $w$ . It is shown that quantum coherence reduces
monotonously with the increase of the acceleration parameter, i.e., the thermal noise induced by Unruh radiation reduces quantum resources. In the limit of $w,r\rightarrow \infty$, the coherence reaches a nonzero asymptotic value $C(\rho_{AB_IC_I})=\frac{\pi}{4}$. In this paper, we call the phenomenon of nonzero asymptotic value for coherence in the limit of infinite acceleration the ``freeze" of coherence.

By tracing over the mode $A$, $B_I$ or $C_I$ from $\rho_{AB_IC_I}$, we further obtain
\begin{eqnarray}\label{GHZ1}
\rho_{B_IC_I}&&=\frac{1}{2}\sum_{m,n=0}^\infty\tanh^{2m}w \tanh^{2n}r \left[\frac{1}{\cosh^2 w\cosh^2 r}|mn\rangle\langle mn|\right.\nonumber \\
&&\left.+\frac{1}{\cosh^4 w\cosh^4 r}(m+1)(n+1)|(m+1)(n+1)\rangle\langle (m+1)(n+1)|\right],
\end{eqnarray}
\begin{eqnarray}\label{GHZ2}
\rho_{AC_I}=\frac{1}{2}\sum_{n=0}^\infty\tanh^{2n}r\left[\frac{1}{\cosh^2 r}|0n\rangle\langle 0n|
+\frac{1}{\cosh^4 r}(n+1)|1(n+1)\rangle\langle 1(n+1)|\right],
\end{eqnarray}
and
\begin{eqnarray}\label{GHZ3}
\rho_{AB_I}=\frac{1}{2}\sum_{m=0}^\infty\tanh^{2m}w\left[\frac{1}{\cosh^2 w}|0m\rangle\langle 0m|
+\frac{1}{\cosh^4 w}(m+1)|1(m+1)\rangle\langle 1(m+1)|\right],
\end{eqnarray}
which suggest that $C(\rho_{B_IC_I})=C(\rho_{AC_I})=C(\rho_{AB_I})=0$. Further calculation suggests that the coherence for any single subsystem is also zero, i.e., $C(\rho_{A})=C(\rho_{B_I})=C(\rho_{C_I})=0$. This means that the coherence of the GHZ state of bosonic fields under Unruh effect is genuinely global, and no coherence exists in any subsystems. Such a property of coherence distribution is correct for any acceleration, including the case of zero acceleration (i.e., $r=w=0$). In other words, the initial tripartite GHZ state of Eq.(\ref{w4}) also has only global coherence. This means that Unruh effect does not disturb the distribution of the GHZ state coherence between different subsystems. Under the influence of Unruh effect, the magnitude of coherence of GHZ state reduces, but its distribution property remains unchanged. According to the discussion of reference \cite{zzzz}, this global coherence is actually a kind of quantum correlation between modes of bosonic fields.

\subsection{Tripartite W state}
\begin{figure}
\includegraphics[scale=1.0]{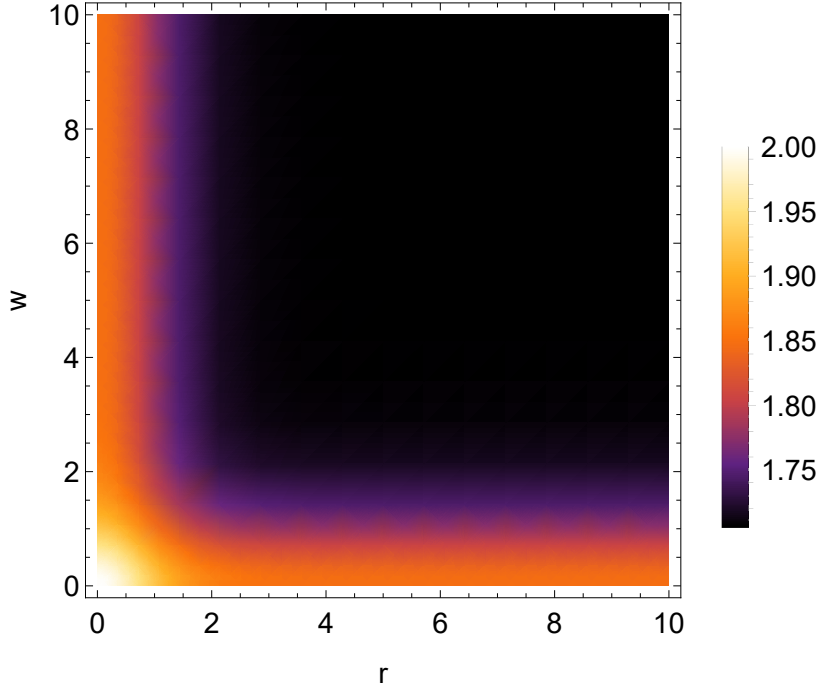}
\caption{Quantum coherence $C(\rho_{AB_IC_I})$ of W state as a function of  the acceleration parameters $r$ and $w$.
 }
\label{Fig2}
\end{figure}

Now we discuss the effect of Unruh radiation on the coherence of tripartite W state of bosonic fields. Assume that Alice, Bob and Charlie share a W state of bosonic fields \cite{S10}
\begin{eqnarray}\label{w8}
|W\rangle=\frac{1}{\sqrt{3}}[|0_{A}0_{B}1_{C}\rangle+|0_{A}1_{B}0_{C}\rangle+|1_{A}0_{B}0_{C}\rangle],
\end{eqnarray}
in the inertial framework. When Bob and Charlie move with acceleration parameters $w$ and $r$ respectively, from the perspective of the accelerated observers, it becomes
\begin{eqnarray}\label{w9}
|W\rangle_{AB_IC_IB_{II}C_{II}}&=&\frac{1}{\sqrt{3}}\sum_{m,n=0}^\infty\tanh^{m}w\tanh^{n}r\left[\frac{1}{\cosh w\cosh^2 r}\sqrt{n+1}|0m(n+1)\rangle|mn\rangle\right. \nonumber \\
&+& \left.
\frac{1}{\cosh^2 w\cosh r}\sqrt{m+1}|0(m+1)n\rangle|mn\rangle
+\frac{1}{\cosh w\cosh r}|1mn\rangle|mn\rangle\right],
\end{eqnarray}
with the notations of field modes as before. After tracing over the unaccessible modes $B_{II}$ and $C_{II}$, we
obtain the reduced density matrix in region $I$ as
\begin{eqnarray}\label{w10}
\rho_{AB_IC_I}&&=\frac{1}{3}\sum_{m,n=0}^\infty\tanh^{2m}w\tanh^{2n}r\left[\frac{1}{\cosh^2 w\cosh^4 r}(n+1)|0m(n+1)\rangle\langle0m(n+1)|\right.\nonumber \\
&&+\frac{1}{\cosh^3 w\cosh^3 r}\sqrt{(m+1)(n+1)}\{|0m(n+1)\rangle\langle0(m+1)n|+h.c.\} \nonumber \\
&&+\frac{1}{\cosh^2 w\cosh^3 r}\sqrt{n+1}\{|0m(n+1)\rangle\langle1mn|+h.c.\}\nonumber \\
&&+\frac{1}{\cosh^4 w\cosh^2 r}(m+1)|0(m+1)n\rangle\langle0(m+1)n|+\frac{1}{\cosh^2 w\cosh^2 r}|1mn\rangle\langle1mn|\nonumber \\
&&+\left.\frac{1}{\cosh^3 w\cosh^2 r}\sqrt{m+1}\{|0(m+1)n\rangle\langle1mn|+h.c.\}\right]
.
\end{eqnarray}
The coherence of this state reads
\begin{eqnarray}\label{w14}
C(\rho_{AB_IC_I})&&=\frac{2}{3}\left[(\frac{1}{\cosh^3 w}\sum_{m=0}^\infty\sqrt{m+1}\tanh^{2m}w)(\frac{1}{\cosh^3 r}\sum_{n=0}^\infty\sqrt{n+1}\tanh^{2n}r)\right. \nonumber\\
&&+\left.\frac{1}{\cosh^3 w}\sum_{m=0}^\infty\sqrt{m+1}\tanh^{2m}w+\frac{1}{\cosh^3 r}\sum_{n=0}^\infty\sqrt{n+1}\tanh^{2n}r\right].
\end{eqnarray}
In the calculations, we have used the relations
\begin{eqnarray}\label{w11}
\frac{1}{\cosh^2 s}\sum_{n=0}^\infty\tanh^{2n}s=1,
\end{eqnarray}
and
\begin{eqnarray}\label{w12}
\frac{1}{\cosh^4 s}\sum_{n=0}^\infty\tanh^{2n}s(n+1)=1.
\end{eqnarray}
Similar to the GHZ state, quantum coherence of W state is also a monotonic decreasing function of acceleration parameters $w$ and $r$ (see Fig.\ref{Fig2}).
In the limit of $w,r\rightarrow\infty$, the coherence has the asymptotic value $C(\rho_{AB_IC_I})=\frac{\pi+4\sqrt{\pi}}{6}$ (freezing phenomenon).

In what follows, we further analyse how the coherence of the W state is distributed in its subsystems under the accelerated motions. After tracing over one of the modes, we obtain the density matrix of bipartite subsystems
\begin{eqnarray}\label{W1}
\rho_{B_IC_I}&&=\frac{1}{3}\sum_{m,n=0}^\infty\tanh^{2m}w\tanh^{2n}r\left[\frac{1}{\cosh^2 w\cosh^2 r}|mn\rangle\langle mn|\right.\nonumber \\
&&+\frac{1}{\cosh^2 w\cosh^4 r}(n+1)|m(n+1)\rangle\langle m(n+1)|\nonumber \\
&&+\frac{1}{\cosh^4 w\cosh^2 r}(m+1)|(m+1)n\rangle\langle (m+1)n|\nonumber \\
&&+\left.\frac{1}{\cosh^3 w\cosh^3 r}\sqrt{(m+1)(n+1)}\{|m(n+1)\rangle\langle (m+1)n|+h.c.\}\right],
\end{eqnarray}
\begin{eqnarray}\label{W2}
\rho_{AC_I}&&=\frac{1}{3}\sum_{n=0}^\infty\tanh^{2n}r\left[\frac{1}{\cosh^2 r}\{|0n\rangle\langle 0n|+|1n\rangle\langle 1n|\}\right.\nonumber \\
&&+\frac{1}{\cosh^4 r}(n+1)|0(n+1)\rangle\langle 0(n+1)|\nonumber \\
&&+\left.\frac{1}{\cosh^3 r}\sqrt{n+1}\{|0(n+1)\rangle\langle 1n|+|1n\rangle\langle 0(n+1)|\}\right],
\end{eqnarray}
and
\begin{eqnarray}\label{W3}
\rho_{AB_I}&&=\frac{1}{3}\sum_{m=0}^\infty\tanh^{2m}w\left[\frac{1}{\cosh^2 w}(|0m\rangle\langle 0m|+|1m\rangle\langle 1m|)\right.\nonumber \\
&&+\frac{1}{\cosh^4 w}(m+1)|0(m+1)\rangle\langle 0(m+1)|\nonumber \\
&&+\left.\frac{1}{\cosh^3 w}\sqrt{m+1}\{|0(m+1)\rangle\langle 1m|+h.c.\}\right].
\end{eqnarray}
The corresponding bipartite coherence reads
\begin{eqnarray}\label{R1}
C(\rho_{B_IC_I})=\frac{2}{3\cosh^3 w\cosh^3 r}(\sum_{m=0}^\infty\sqrt{m+1}\tanh^{2m}w)(\sum_{n=0}^\infty\sqrt{n+1}\tanh^{2n}r),
\end{eqnarray}
\begin{eqnarray}\label{R2}
C(\rho_{AC_I})=\frac{2}{3\cosh^3 r}\sum_{n=0}^\infty\sqrt{n+1}\tanh^{2n}r,
\end{eqnarray}
and
\begin{eqnarray}\label{R3}
C(\rho_{AB_I})=\frac{2}{3\cosh^3 w}\sum_{m=0}^\infty\sqrt{m+1}\tanh^{2m}w.
\end{eqnarray}
Further calculation suggests that the monomeric coherence of any single subsystems is zero, i.e., $C(\rho_{A})=C(\rho_{B_I})=C(\rho_{C_I})=0$.

\begin{figure}
\begin{minipage}[t]{0.43\linewidth}
\centering
\includegraphics[width=2.7in,height=5.2cm]{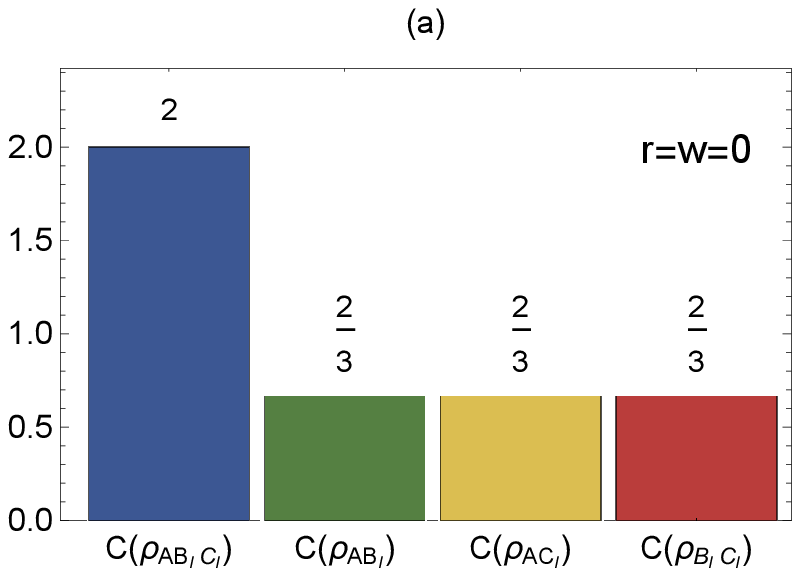}
\label{fig3a}
\end{minipage}%
\begin{minipage}[t]{0.43\linewidth}
\centering
\includegraphics[width=2.7in,height=5.2cm]{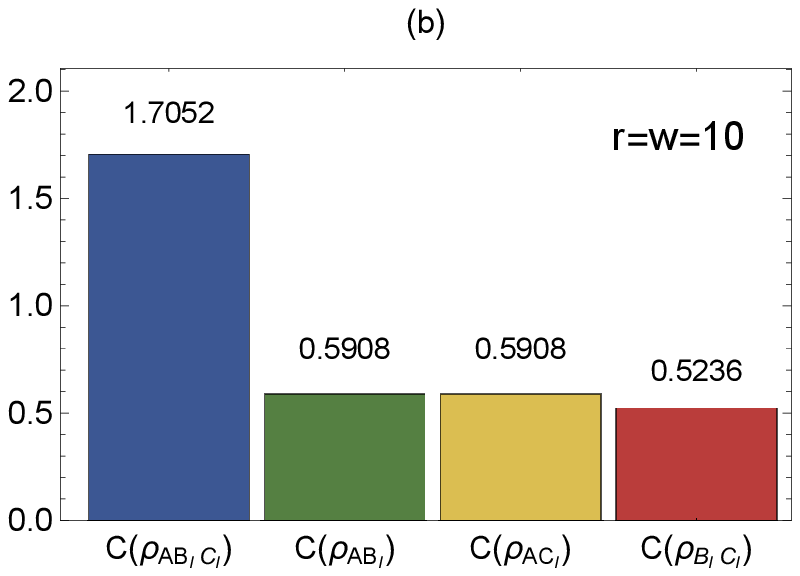}
\label{fig3b}
\end{minipage}%

\vspace{1cm}
\begin{minipage}[t]{0.43\linewidth}
\centering
\includegraphics[width=2.7in,height=5.2cm]{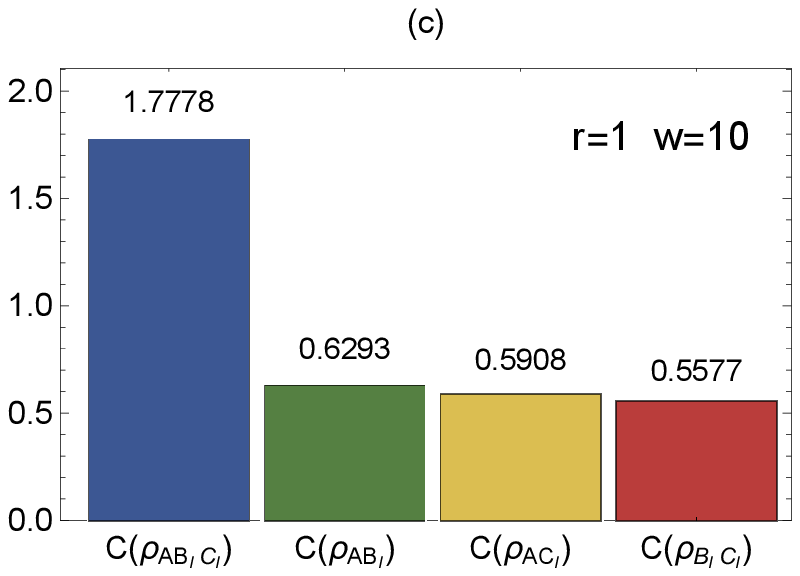}
\label{fig3c}
\end{minipage}%
\begin{minipage}[t]{0.43\linewidth}
\centering
\includegraphics[width=2.7in,height=5.2cm]{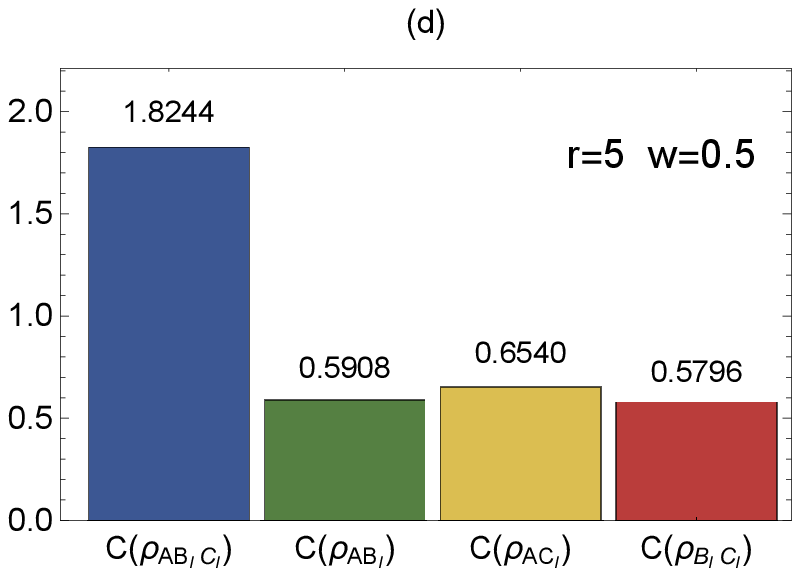}
\label{fig3d}
\end{minipage}%
\caption{Values of $C(\rho_{AB_IC_I})$, $C(\rho_{AB_I})$, $C(\rho_{AC_I})$ and $C(\rho_{B_IC_I})$ for W state under different sets of acceleration parameters. The coherence distribution relationship of Eq.(\ref{DIS}) is always fulfilled. }
\label{Fig3}
\end{figure}

After a careful inspection of Eq.(\ref{w14}) and Eqs.(\ref{R1})-(\ref{R3}), we find
a distribution relation for the coherence of the tripartite W state
\begin{eqnarray}\label{DIS}
C(\rho_{B_IC_I})+C(\rho_{AC_I})+C(\rho_{AB_I})=C(\rho_{AB_IC_I}).
\end{eqnarray}
The above calculations suggest that the coherence of the evolved W state is essentially existed in a style of bipartite forms, and the total coherence of the whole system is equal to the sum of all the genuine bipartite types of coherence. Such a property of coherence distribution is also correct for zero acceleration, meaning that the coherence distribution of tripartite W state is robust to Unruh effect. Under the influence of Unruh effect, the magnitudes of all the bipartite types of coherence reduce continually, but the coherence distribution remains unchanged, i.e. the relationship Eq.(\ref{DIS}) is fulfilled for any acceleration. In Fig.\ref{Fig3}, we numerically show this distribution relationship under different sets of acceleration parameters : (a) $r=w=0$; (b) $r=w=10$; (c) $r=1$ and $w=10$; (d) $r=5$ and $w=0.5$. For clarity, we mark the values of coherence on the top of the corresponding column charts.

\section{Extension to $N$-partite systems}
In this section, we want to extend the above investigations from tripartite systems to the arbitrary $N$-partite systems ($N\geq 3$). Assume that there is a $N$-partite GHZ state,
\begin{eqnarray}\label{N1}
GHZ_{123...N}=\frac{1}{\sqrt{2}}( |00...00\rangle+|11...11\rangle),
\end{eqnarray}
or W state
\begin{eqnarray}\label{N2}
W_{123...N}=\frac{1}{\sqrt{N}}( |10...00\rangle+|01...00\rangle+...+|00...01\rangle),
\end{eqnarray}
of the bosonic modes in inertial frame. Where the mode $i$ ($i=1,2,...,N$) is detected by observer $O_{i}$. Now we assume that $n$ ($2\leq n\leq N-1$)
observers move with uniform accelerations, and the rest $N-n$ observers stay still in the inertial frame.
After tedious but straightforward calculations, we acquire the expressions for the coherence of $N$-partite GHZ and W states in noninertial frames, \begin{eqnarray}\label{NGHZ}
 C(GHZ)=\prod_{i=1}^n(\frac{1}{\cosh^{3}r_i}\sum_{m=0}^\infty\sqrt{m+1}\tanh^{2m}r_i),
\end{eqnarray}
\begin{eqnarray}\label{NGHW}
C(W)&&=\frac{2}{N}\left[\sum_{i,j=1 (i\neq j)}^n(\frac{1}{\cosh^{3}r_i}\sum_{z=0}^\infty\sqrt{z+1}\tanh^{2z}r_i)
(\frac{1}{\cosh^{3}r_j}\sum_{m=0}^\infty\sqrt{m+1}\tanh^{2m}r_j)\right.\nonumber\\
&&+\left.(N-n)\sum_{i=1}^n(\frac{1}{\cosh^{3}r_i}\sum_{m=0}^\infty\sqrt{m+1}\tanh^{2m}r_i)
+\frac{(N-n)(N-n-1)}{2}\right],
\end{eqnarray}
where the acceleration parameter $r_i$ ($i=1,2,...,n$) corresponds to the accelerated
observer $O_{i}$. Obviously, the coherence of both $N$-partite GHZ and W states depends on the number $n$ of the accelerated observers, but the coherence of $N$-partite GHZ state is independent of the number $N$ of the subsystems.
\begin{figure}
\begin{minipage}[t]{0.43\linewidth}
\centering
\includegraphics[width=2.7in,height=5.2cm]{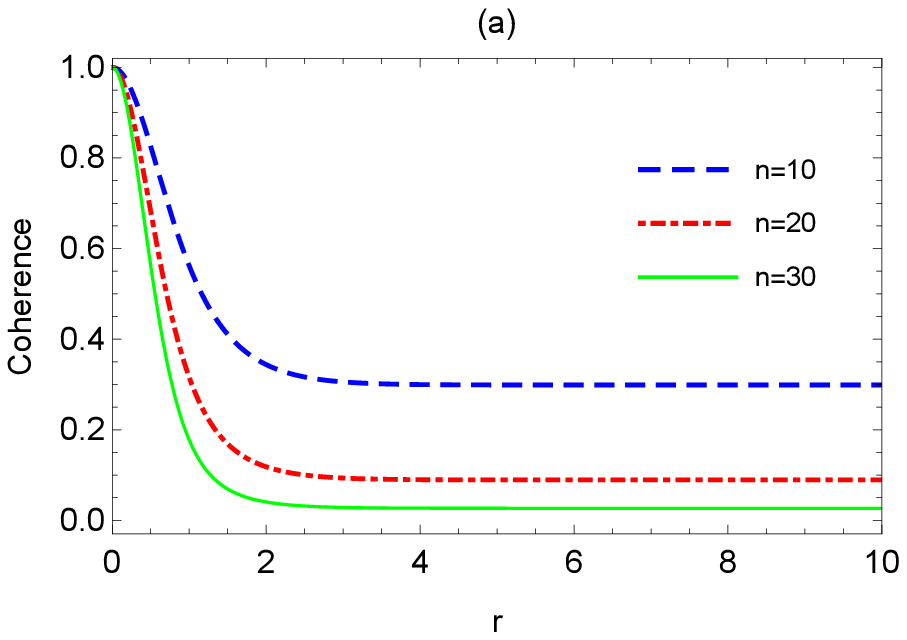}
\label{fig4a}
\end{minipage}%
\begin{minipage}[t]{0.43\linewidth}
\centering
\includegraphics[width=2.7in,height=5.2cm]{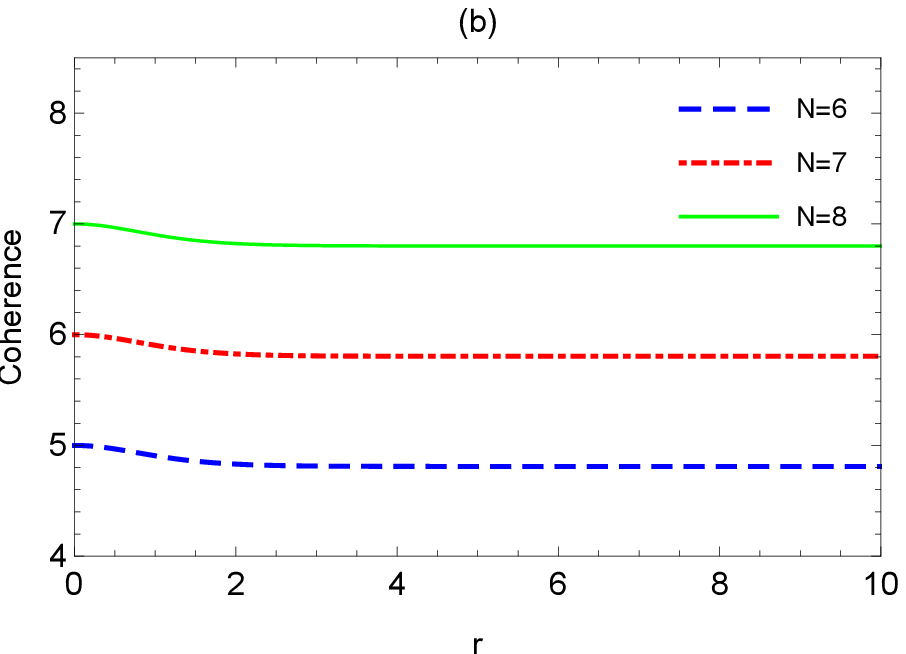}
\label{fig4b}
\end{minipage}%

\begin{minipage}[t]{0.43\linewidth}
\centering
\includegraphics[width=2.7in,height=5.2cm]{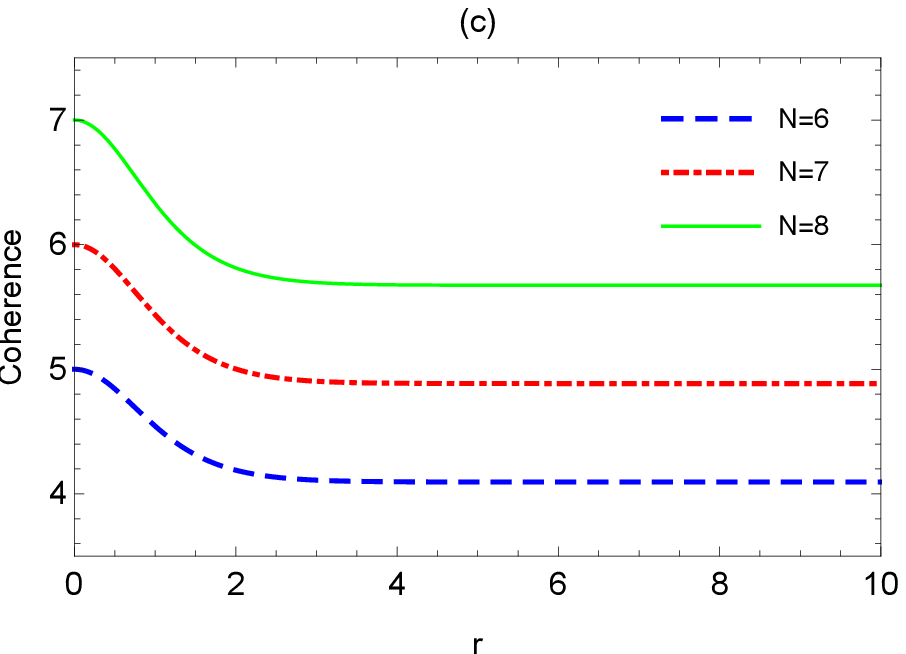}
\label{fig4c}
\end{minipage}%
\begin{minipage}[t]{0.43\linewidth}
\centering
\includegraphics[width=2.7in,height=5.2cm]{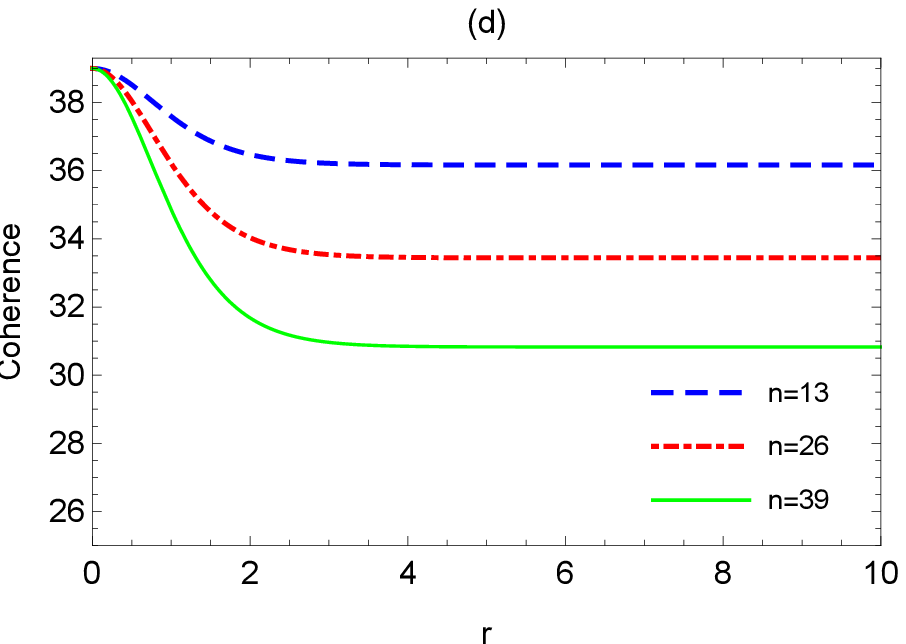}
\label{fig4d}
\end{minipage}%
\caption{Coherence as a function of the acceleration parameter $r$, (a) for $N$-partite GHZ state and (b)-(d) for $N$-partite W state. In (b), we assume one observer accelerates and the rest $N-1$ are in the inertial frames; while in (c), it is assumed one observer is in the inertial frame and the rest $N-1$ observers accelerate. In (d), we set the number of the subsystems $N=40$. In the case of many accelerated observers, we set the accelerated parameters are the same.}
\label{Fig4}
\end{figure}

In Fig.\ref{Fig4}(a), we plot the coherence of $N$-partite GHZ state as the function of the acceleration parameter $r$ for different number $n$ of the accelerated observers. For simplicity, we set the acceleration parameters of the $n$ accelerated observers the same (denoted by $r$). We see that the coherence reduces with acceleration. In the limit of infinite acceleration $r\rightarrow \infty$, the coherence approaches to the asymptotic value $(\sqrt{\pi/4})^{n}$. From the three lines given in the subgraph, we find that the coherence is also a decreasing function of $n$.

In Fig.\ref{Fig4}(b)-(d),  we plot the coherence of $N$-partite W state as the function of the accelerated observers $r$ for different $n$ and $N$. We see that generally the coherence is a decreasing function of the acceleration parameter $r$. Compared \ref{Fig4}(b)-(d), we find that the coherence increases with the number $N$ of the subsystems and decreases with the number $n$ of the accelerated observers.

Next, we study the distribution of coherence for the $N$-partite systems. For the N-partite GHZ state,
we still find that the quantum coherence is genuinely global, and no coherence exists in any subsystems even if under the non-inertial frames. For $N$-partite W state, the coherence is essentially bipartite types and the following distribution relationship follows
\begin{equation}\label{SUM}
\sum_{i,j=1(i\neq j)}^{n}C(\rho_{I_iI_j})+\sum_{i=1}^{n}\sum_{j=1}^{N-n}C(\rho_{I_iX_j})
+\sum_{i,j=1(i\neq j)}^{N-n}C(\rho_{X_iX_j})= C(W),
\end{equation}
where $C(\rho_{I_iI_j})=\frac{2}{N}(\frac{1}{\cosh^{3}r_i}\sum_{z=0}^\infty\sqrt{z+1}\tanh^{2z}r_i)
(\frac{1}{\cosh^{3}r_j}\sum_{m=0}^\infty\sqrt{m+1}\tanh^{2m}r_j)$
denotes the bipartite coherence for the modes corresponding to accelerated observers $I_i$ and $I_j$, and $C(\rho_{I_iX_j})=\frac{2}{N}\frac{1}{\cosh^{3}r_i}\sum_{m=0}^\infty\sqrt{m+1}\tanh^{2m}r_i$
denotes the bipartite coherence for the modes corresponding to the accelerated observer $I_i$ and the inertial observer
$X_j$, and $C(\rho_{X_iX_j})=\frac{2}{N}$ denotes the bipartite coherence for the modes corresponding to the two inertial observers $X_i$ and $X_j$. Eq. (\ref{SUM}) means that the total coherence of the $N$-partite W state is equal to the sum of all the bipartite types of coherence, under the acceleration of arbitrary number of observers.

\section{Conclusions}
In this paper, the acceleration effect with two accelerated observers on the coherence of tripartite GHZ and W states for bosonic fields have been investigated. It has been shown that the coherence for both tripartite GHZ and W states reduces with acceleration of the observers, and approaches to a nonzero asymptotic value in the limit of infinite acceleration (freezing of coherence). The coherence of the tripartite GHZ state is always global under arbitrary accelerations, but the coherence of the tripartite W state is essentially bipartite types and the total coherence of the tripartite W state is equal to the sum of all the bipartite types of coherence.

We have also extended the investigations from tripartite systems to the $N$-partite systems and found, apart from the similar properties for the tripartite systems, that the degradation of coherence for the $N$-partite systems depends on the number of accelerated observers: The more the number of the accelerated observers is, the faster the coherence reduces. In the limit of infinite acceleration, the coherence approaches to a nonzero value that depends on the number of the accelerated observers. Interestingly, the coherence of $N$-partite GHZ state depends only on the number of the accelerated observers and not on the total number $N$, while the coherence of the $N$-partite W state depends on both of them. It has been also found that the coherence of $N$-partite GHZ states is global, but the coherence of $N$-partite W states is essentially bipartite types and the total coherence is equal to the sum of all the bipartite types of coherence.

It is worthwhile to point out that we have assumed more than two observers accelerate in the description of coherence, for both tripartite and $N$-partite systems. However, the results also apply to the situation when only one observer accelerates. This can be inferred by keeping one acceleration parameter nonzero and setting all the rest acceleration parameters to be zero. It is easily to imagine that the distribution characters of coherence remain unchanged, except for the change of expressions for the relevant coherence.

Coherence and entanglement are related to each other closely \cite{zzzz}, both of them are the important resources in quantum information processing. However, under the influence of Unruh effect, these two resources also have different behaviors. For example, the bosonic coherence of both the tripartite GHZ and W states reduces more slowly compared with the case of fermionic fields\cite{S1,SQW}, with the increasing of the observer's accelerations. In contrast, the fermionic entanglement is more robust against the Unruh effect than the case of bosonic fields\cite{Q12,Q14,Q17}. In addition, it has been found that the bosonic entanglement vanishes when the acceleration approaches to infinite \cite{Q14,Adesso2007,Schuller-Mann,Richter2015}. However, we have found in this paper that the bosonic coherence of both GHZ and W states freezes to nonzero value in the infinite acceleration limit. We hope our results can further enrich the research about relativistic quantum resources and provide helps for some quantum information tasks.

In the end, we point out that the properties about the degradation and distribution of quantum coherence induced by Unruh effect can be applied to the case of some gravitational fields, such as the Schwarzschild black hole. In fact, similar analogy on the effect of quantum entanglement, for both bosonic and fermionic fields, have been carried out already\cite{Schuller-Mann,Richter2015}. In this analogy, the inertial observers
in the Minkowski spacetime correspond to the free-falling
observers in the Schwarzschild spacetime and the uniformly accelerated observers correspond to the static observers outside the black hole. Each observer measures his own mode and sends the result to an appointed adjudicator who computes the coherence through the received results.

\begin{acknowledgments}
This work is supported by the National Natural
Science Foundation of China (Grant No. 11275064), and the Construct Program of the National Key Discipline.	
\end{acknowledgments}


\end{document}